\documentclass[twocolumn,aps]{revtex4}
\usepackage{amsmath}
\usepackage{amssymb}
\usepackage{graphicx}
\usepackage{dcolumn}
\usepackage{bm}
\usepackage{xcolor}

\begin{document}

\title{Linking boundary conditions for kinetic and hydrodynamic description of fermion gas}

\author{O. E. Raichev}

\affiliation{Institute of Semiconductor Physics NAS of Ukraine, Prospekt Nauki 41, 03028 Kyiv, Ukraine}

\date{\today}

\begin{abstract}

An approximate analytical solution of the boundary slip problem in magnetic field 
is obtained by using the general form of boundary conditions for the distribution 
function of fermions with the isotropic energy spectrum. Exact numerical calculations 
of the slip length for different models of angle-dependent specularity parameter and 
application of the results to the description of the Poiseuille flow demonstrate the 
reliability of the approximate solution for establishing a direct link between the 
hydrodynamic and the kinetic approaches to transport in bounded fermion systems.
 
\end{abstract}

\maketitle

Studies of hydrodynamic phenomena in electron transport in solids \cite{gurzhi}-\cite{gupta} 
are most often based on investigation of current flow in the samples of small size, 
where the presence of boundaries considerably influences the transport properties. 
In particular, the non-ideal boundaries, where electrons change their momenta as a 
result of non-specular reflection, are responsible for drift velocity gradients 
affecting the hydrodynamic flow via the viscosity. The hydrodynamic regime occurs 
when the rate of electron-electron collisions conserving the local momentum of electron 
system becomes larger than the rates of momentum-changing scattering of electrons by 
impurities, phonons, and boundaries. In this case, the transport is described in terms 
of the coordinate-dependent drift velocity ${\bf u}({\bf r})$, or the related electric 
current ${\bf j}=en{\bf u}$, where $e$ and $n$ are the electron charge and density, 
governed by the Navier-Stokes equation (NSE). Beyond the hydrodynamic regime, one has 
to use a more general albeit more complicated approach based on solution of the Boltzmann 
kinetic equation for the distribution function $f_{\bf p}({\bf r})$ that depends on both 
the momentum ${\bf p}$ and coordinate ${\bf r}$. The kinetic equation approach covers 
all classical transport regimes and is indispensable for description of the transition 
between the quasi-ballistic and hydrodynamic regimes, when the modifications of 
electrical resistance dependence on the temperature and magnetic field serve as 
hallmarks for the onset of hydrodynamic behavior \cite{scaffidi,alekseev2,holder,
sulpizio,raichev1,gupta}. 

Establishing a connection between the Boltzmann equation and the NSE in bounded systems
is of particular importance. The NSE itself can be derived from the Boltzmann equation 
either by applying the method of moments \cite{grad} or, equivalently, by expanding 
$f_{\bf p}({\bf r})$ into series of harmonics of the angle of ${\bf p}$, and neglecting 
the higher-order moments under the restrictions imposed by the hydrodynamic regime. This 
commonly accepted procedure is straightforward because the relation between the current 
and distribution function is direct and assumes just an integration over the 
entire momentum space. However, a connection between the boundary conditions (BC) 
used in the kinetic theory and hydrodynamic BC is not so obvious because of the 
different mathematical nature of these conditions. The hydrodynamic BC relates the 
tangential component of ${\bf u}$ with its derivative at the boundary. The kinetic 
BC relates the distribution functions from {\it different} momentum semi-spaces 
corresponding to incident and reflected particles. Furthermore, the boundary 
reflection mixes different angular harmonics of $f_{\bf p}({\bf r})$, so  
the use of truncated sets of moments generally fails near the boundary. 

The general form of the hydrodynamic BC for both three- (3D) and 
two-dimensional (2D) systems is 
\begin{eqnarray}
{\bf u}_t= l_S \nabla_n {\bf u}_t,
\end{eqnarray}
where ${\bf u}_t$ is the tangential component of the drift velocity, $\nabla_n {\bf u}_t$ 
is its normal inward derivative, and $l_S$ is the slip length that plays important role in 
hydrodynamics. The boundary slip problem, i.e., determination of the relation of $l_S$ to the 
kinetic properties of gas or liquid became a subject of study since the foundation of the kinetic 
theory \cite{maxwell}. For 3D gases, approximate results relating $l_S$ to the viscosity 
have been obtained, and the exact values of $l_S$ have been calculated in the limit 
of fully diffuse boundary reflection \cite{einzel}. The increase in $l_S$ with increasing 
degree of specularity has been also discussed \cite{maxwell,jensen,einzel,pellegrino2}. 
Until recently, the existing results in the boundary slip problem were ignored in the 
rapidly developing hydrodynamics of fermion gas in solids, and $l_S$ was often 
considered as a freely adjustable parameter. The authors of Ref. \cite{kiselev} have drawn 
attention to the problem and calculated $l_S$ for 2D fermions in the cases of fully 
diffuse and nearly specular reflection. Despite these achievements, a full
and clear correspondence between the kinetic and hydrodynamic BC is still missing. 

In this Letter, an approximate expression for the slip length is derived by using the general 
form of BC for the distribution function and in the presence of magnetic field. Next, the 
exact slip length is calculated for nonzero specularity of boundary reflection, including 
different models of angle-dependent specularity parameter. It is found that the exact 
results approach the approximate ones when specularity increases. This important 
property gives more reliability to the approximate results and justifies their 
application in the hydrodynamics of fermion systems, as further demonstrated by 
comparing hydrodynamic and kinetic solutions of the transport problem in narrow 
2D channels. 
 
First, a brief review of the BC used in the kinetic theory is presented. 
It is assumed below that the particles are described by the isotropic energy spectrum 
$\varepsilon=\varepsilon_p$ and the boundary scattering is elastic, $|{\bf p}|=|{\bf p}'|=
p_{\varepsilon}$. The BC for the distribution function at the hard-wall boundary in 3D 
media is
\begin{eqnarray} 
f^{+}_{\bf p}=f^{-}_{\bf p} + \int_+ \frac{d \Omega'}{4 \pi} c'_n P_{\varepsilon}(
{\bf p},{\bf p}') (f^-_{{\bf p}'}-f^-_{\bf p}),
\end{eqnarray}
where $f^{\pm}_{\bf p}=f_{{\bf p}_t,\pm p_n}$ are the distribution functions of reflected ($+$) 
and incident ($-$) particles, $p_n=p_{\varepsilon} \sin \varphi$ and ${\bf p}_t$ 
are the normal and tangential components of the momentum, ${\bf c}={\bf p}/|{\bf p}|$ is the unit 
vector along the momentum, and $\int_+ \frac{d \Omega'}{4 \pi} ...$ 
denotes averaging over the solid angle of ${\bf p}'$ in the $+$ hemisphere, where $\sin \varphi'>0$. 
Equation (2) is a general form of the integral relation between the distribution functions of 
incident and reflected particles guaranteeing zero particle flow through the boundary for arbitrary 
$f_{\bf p}$. This equation can be derived from the quantum-mechanical reflection problem, see \cite{soffer}, 
\cite{falkovski}, \cite{okulov}, and Sec. 44 in \cite{VRbook}. 
The boundary scattering is described by the function $P_{\varepsilon}$, which is symmetric 
with respect to permutation of momenta, $P_{\varepsilon}({\bf p},{\bf p}')=
P_{\varepsilon}({\bf p}',{\bf p})$,
and equal to zero at $\varphi=0$ as the reflection is specular at grazing incidence. For macroscopically 
isotropic boundary, $P_{\varepsilon}({\bf p},{\bf p}')$ is invariant with respect to simultaneous rotation 
of ${\bf p}_t$ and ${\bf p}'_t$. The probability of specular reflection at the angle $\varphi$ is 
given by the specularity parameter ${\rm r}_{\varepsilon \varphi}=1-\int_+ \frac{d \Omega'}{4 \pi} c'_n 
P_{\varepsilon}({\bf p},{\bf p}')$. In the case of non-correlated boundary scattering, when the 
non-specular reflection (whose probability is $1-{\rm r}_{\varepsilon \varphi}$) is isotropic, 
Eq. (2) is simplified to the form 
\begin{eqnarray} 
f^{+}_{\bf p}={\rm r}_{\varepsilon \varphi} f^-_{\bf p}+
(1-{\rm r}_{\varepsilon \varphi}){\overline f}_{\varepsilon}, \nonumber \\ 
{\overline f}_{\varepsilon}= \left. \int_+ \frac{d \Omega}{4 \pi} c_n 
(1-{\rm r}_{\varepsilon \varphi}) f^-_{\bf p} \right/ 
\int_+ \frac{d \Omega}{4 \pi} c_n (1-{\rm r}_{\varepsilon \varphi}).
\end{eqnarray} 
The limiting cases ${\rm r}_{\varepsilon \varphi}=1$ and ${\rm r}_{\varepsilon \varphi}=0$ 
describe fully specular and fully diffuse reflection. In spite of less generality compared 
to Eq. (2), Eq. (3) is far more convenient for applications because it is not 
an integral relation, and the boundary properties can be modelled by specifying 
the magnitude and the angular dependence of ${\rm r}_{\varepsilon \varphi}$. 

The current-penetrable boundary can be described by the in-flow BC \cite{guo}: 
$f^{+}_{\bf p}= {\cal G}_{\varepsilon}(\varphi)$, where ${\cal G}$ models inward 
emission of particles. This BC can be viewed as a particular case 
of Eq. (3) in the fully diffuse limit.

Equations (2) and (3) are adopted for 2D systems by reducing 
${\bf p}_t$ to the scalar variable $p_{\varepsilon} \cos \varphi$ and by substituting 
$\int_+ \frac{d \Omega'}{4 \pi} \ldots \rightarrow \int_0^{\pi} \frac{d \varphi'}{2 \pi} \ldots$. 
The function $P_{\varepsilon}({\bf p},{\bf p}')$ can be written as 
$P_{\varepsilon}(\varphi, \varphi')$.

Consider now a steady-state linear-transport classical kinetic problem for charged fermions 
in a homogeneous magnetic field ${\bf B}$ directed parallel to the boundary. The distribution 
function is presented as  
\begin{eqnarray}
f_{\bf p}({\bf r})=f_{\varepsilon} + \delta f_{\bf p}({\bf r})=  
f_{\varepsilon} - \frac{ \partial f_{\varepsilon}}{\partial 
\varepsilon} [g_{{\bf p}}({\bf r})- e\Phi({\bf r})],
\end{eqnarray}
where $f_{\varepsilon}$ is the equilibrium Fermi-Dirac distribution and $\Phi({\bf r})$ 
is the electrostatic potential. The BC given by Eqs. (2) and (3) are valid as well for the 
non-equilibrium correction $g_{\bf p}$. In the hydrodynamic regime, 
\begin{eqnarray}
g_{{\bf p}}=g_0+{\bf g}_{\varepsilon} \cdot {\bf c} + Q_{\varepsilon}^{
\alpha \beta}(c_{\alpha}c_{\beta}-\delta_{\alpha \beta}/d),
\end{eqnarray}
where $d$ is the dimensionality of the system, $g_0({\bf r})=eV({\bf r})$, $V$ is the 
non-equilibrium electrochemical potential, ${\bf g}_{\varepsilon}({\bf r})$ describes
the drift velocity ${\bf u}=\langle {\bf g}_{\varepsilon}/p_{\varepsilon} \rangle$, and 
$Q_{\varepsilon}^{\alpha \beta}({\bf r})$ is the symmetric tensor describing the momentum 
flow density $\Pi_{\alpha \beta}=2n \langle Q_{\varepsilon}^{\alpha \beta} \rangle/(d+2)$. 
The average over energy is $\left< A_{\varepsilon} \right> \equiv \int d \varepsilon D_{\varepsilon} 
v_{\varepsilon} p_{\varepsilon} (-\partial f_{\varepsilon}/\partial \varepsilon ) 
A_{\varepsilon}/dn$, where $D_{\varepsilon}$ is the density of states and $v_{\varepsilon}$ 
is the absolute value of the group velocity. Equation (5) is a truncated 
expansion of the exact $g_{{\bf p}}$ in powers of $c_{\alpha}$. Beyond the hydrodynamic regime, 
it is necessary to include all higher-order terms in this expansion.

It is specified below that a flat boundary is placed at $y=0$ and the magnetic field is 
directed along the $Oz$ axis. Evaluating the collision integral in the kinetic equation in the 
elastic relaxation-time approximation, and taking into account that the drift velocity depends 
only on $y$, one can express the components of $Q_{\varepsilon}^{\alpha \beta}$ (and hence of 
the viscous stress tensor $-\Pi_{\alpha \beta}$) as follows:
\begin{eqnarray}
Q_{\varepsilon}^{xx}=-Q_{\varepsilon}^{yy}= l_{\bot} \omega_c \tau \nabla_y g^x,
~ Q_{\varepsilon}^{xy}=-\frac{l_{\bot}}{2} \nabla_y g^x, \nonumber \\
Q_{\varepsilon}^{xz}= \frac{l_{||}}{2} \omega_c \tau \nabla_y g^z,~ Q_{\varepsilon}^{yz}=
-\frac{l_{||}}{2} \nabla_y g^z,~ Q_{\varepsilon}^{zz}=0,
\end{eqnarray}
where $\omega_c=-eB/mc$ is the cyclotron frequency and $m=p_{\varepsilon}/v_{\varepsilon}$ 
is the effective mass. The lengths
$l_{\bot}=l_{\varepsilon}/[1+(2 \omega_c \tau)^2]$ and $l_{||}=l_{\varepsilon}/[1+(\omega_c 
\tau)^2]$, characterize the stress for the drift in the directions transverse and longitudinal 
with respect to ${\bf B}$. The stress at $B=0$ is determined by $l_{\varepsilon}=v_{\varepsilon} 
\tau$, where $1/\tau=1/\tau_{2}+1/\tau_{e}$ 
\cite{alekseev1} is the relaxation rate of the second angular harmonic of the 
distribution function. This rate is a sum of the contributions from the momentum-changing scattering and 
from the momentum-conserving scattering between the particles, characterized 
by the times $\tau_{2}$ and $\tau_e$, respectively. In the hydrodynamic regime, $\tau_2 \gg \tau_{e}$, so $l_{\varepsilon} \simeq l_e = v_{\varepsilon} \tau_{e}$. Applying Eqs. (5) and 
(6) together with the hydrodynamic expression ${\bf g}_{\varepsilon}({\bf r})=p_{\varepsilon} {\bf u}(
{\bf r})$ into the kinetic equation, multiplying the latter by ${\bf p}$, and integrating 
over momentum, one gets the linearized NSE
\begin{eqnarray}
n^{-1} \eta_{\alpha \beta} \nabla_y^2 u_{\beta} + (eB/c) \epsilon_{z \alpha \beta} u_{\beta} 
- \zeta u_{\alpha} = e\nabla_{\alpha} V,
\end{eqnarray}
which is valid both for 3D media and for 2D layers in the $xy$ plane. Here, $\epsilon_{z \alpha 
\beta}$ is the antisymmetric unit tensor, $\zeta=\langle m/\tau_1\rangle$, 
$\tau_1$ is the relaxation time of the first angular harmonic of the distribution function, 
also known as the transport time, and $\eta_{\alpha \beta}$ is the dynamic viscosity tensor. 
The components of $\eta_{\alpha \beta}$ contributing to Eq. (7) are transverse, longitudinal, 
and Hall viscosities:
\begin{eqnarray}
\eta_{xx}=n\frac{\langle p_{\varepsilon} l_{\bot} \rangle}{d+2},~ \eta_{zz}=n \frac{\langle 
p_{\varepsilon} l_{||} \rangle}{d+2},~ \eta_{yx}= 2 n\frac{\langle \omega_c \tau p_{\varepsilon} 
l_{\bot} \rangle}{d+2}.
\end{eqnarray}

The BC for ${\bf u}$ in Eq. (7) can be derived on an equal footing, by using Eqs. (5) and (6). 
After expressing $g_{\bf p}$ in terms of $g^{\pm}_{\bf p}=g_{{\bf p}_t,\pm p_n}$, the tangential 
momentum flow density in the direction normal to the boundary is 
\begin{eqnarray}
\Pi_{\alpha y}({\bf r}) = n d \left< \int_+ \frac{d \Omega}{4 \pi} c_{\alpha} c_y 
[g^+_{\bf p}({\bf r})-g^-_{\bf p}({\bf r})] \right>, 
\end{eqnarray}
where $\alpha=x,z$ for 3D and $\alpha=x$ for 2D systems. Far from the boundary, Eq. (9) 
is an identity, since it is reduced to the definition of $\Pi_{\alpha y}$ as the 
integral of the product $p_{\varepsilon} v_{\varepsilon} c_{\alpha} c_{y} f_{\bf p}$ over the 
momentum space. Further application of Eq. (9) at the boundary $y=0$ is based on Maxwell's 
idea \cite{maxwell}. It is assumed that the hydrodynamic form (5), which is valid away 
from the boundary, remains valid everywhere for the {\it incident} ($-$) particles. 
Expressing $g^+_{\bf p}$ in Eq. (9) through $g^-_{\bf p}$ with the aid of the BC (3) and 
then using Eq. (5) for $g^-_{\bf p}$, one finally obtains two terms proportional to 
$\langle g_{\varepsilon}^{\alpha} \rangle$ and $\langle Q_{\varepsilon}^{\alpha y} \rangle$. 
After applying Eq. (6) together with ${\bf g}=p_{\varepsilon} {\bf u}$, Eq. (9) assumes the 
form of Eq. (1), where $l_S$ is expressed as 
\begin{eqnarray}
l_S^{(i)}=  \frac{\left< p_{\varepsilon} l_{i} \lambda_{2}^+ \right>}{\left< p_{\varepsilon} 
\lambda_{1}^-\right>}, \lambda_{k}^{\pm}=\int_0^{\pi/2} \! \! \! d \varphi (1 \pm {\rm r}_{\varepsilon \varphi}) 
\cos^d \varphi \sin^k \varphi,
\end{eqnarray}
where $i=\bot,||$ for $d=3$ and $i=\bot$ for $d=2$. Application of the general BC (2) 
instead of BC (3) is reduced to the formal substitution ${\rm r}_{\varepsilon \varphi} 
\rightarrow \tilde{{\rm r}}_{\varepsilon \varphi}$ in Eq. (10), where 
\begin{eqnarray}
\tilde{{\rm r}}_{\varepsilon \varphi} = 1-\int_+ \frac{d \Omega'}{4 \pi} \sin \varphi'
P_{\varepsilon}({\bf p},{\bf p}') \left[1- \frac{\cos \varphi'}{\cos \varphi} 
\cos \theta_- \right]
\end{eqnarray}
and $\theta_-$ is the difference of the azimuthal angles of ${\bf p}$ and ${\bf p}'$. 
In the 2D case $\tilde{{\rm r}}_{\varepsilon \varphi} = 1-\int_0^{\pi}\frac{d \varphi'}{2 \pi} 
\sin \varphi' P_{\varepsilon}(\varphi,\varphi') [1- \cos \varphi'/\cos \varphi]$. 
 
If energy dependence of ${\rm r}_{\varepsilon \varphi}$ is absent or inessential, Eq. (10) 
is rewritten as
\begin{eqnarray}
l_S^{(i)}= \Lambda_d \ell_i ,~~\ell_i=(d+2)\eta_{i}/\langle p_{\varepsilon} \rangle n,~~\Lambda_{d} = 
\lambda_{2}^{+}/\lambda_{1}^{-},
\end{eqnarray} 
where $\eta_{\bot} \equiv \eta_{xx}$ and $\eta_{||} \equiv \eta_{zz}$, see Eq. (8). The
slip length is equal to the product of the slip coefficient $\Lambda_d$ by the 
length $\ell_i$ related to the diagonal components of the viscosity. The relation of $l_S$ to 
viscosity, known previously at $B=0$, persists at finite $B$, when 
the viscosity is $B$-dependent. Equation (12) is applicable, e.g., if the specularity 
parameter is a constant ${\rm r}$. Then the slip coefficients are $\Lambda_{3}=
(8/15)(1+{\rm r})/(1-{\rm r})$ \cite{jensen} and $\Lambda_{2}=(3 \pi/16)(1+{\rm r})/(1-{\rm r})$. 
Next, Eq. (12) is always applicable to degenerate fermion gases, as the energy average 
fixes $\varepsilon$ at the Fermi level $\varepsilon_F$. Eqs. (10)-(12) are derived 
in the hydrodynamic regime, $l \ll l_1$, where $l=\left<l_{\varepsilon} \right>$ and 
$l_1=\left< v_{\varepsilon} \tau_1 \right>$ is the transport mean free path length. 
They apply for fermions with arbitrary energy spectrum and for ordinary gases if 
$f_{\varepsilon}$ is the Boltzmann distribution.

The subsequent consideration contains a more detailed calculation of $l_S$ and applications 
of the results to a transport problem. The case of degenerate fermion gas is studied, so 
$\ell_i=l_i$ at $B \neq 0$ and $\ell_i=l$ at $B=0$. Since $\varepsilon=\varepsilon_F$, the 
energy index is omitted ($l_{\varepsilon}=l$, $v_{\varepsilon}=v$, etc.) here and below.

The results given by Eqs. (10)-(12) are approximate because the distribution function of 
incident particles loses its hydrodynamic form in the narrow Knudsen layer near 
the boundary \cite{einzel}. To improve the accuracy and to find the exact slip length, both numerical and 
analytical methods have been developed for 3D gases \cite{jensen,welander,willis,albertoni,einzel2,
latyshev,gu}. The problem of {\it exact} slip length is reduced to solution of an integral equation for 
the drift velocity near the boundary. Below, this problem is solved at $B=0$, by exploiting the integral 
equation for the current density ${\bf j}=en{\bf u}$ in a 2D channel obtained \cite{dejong} 
from the Boltzmann equation with application of the BC (3), see also Ref. \cite{raichev1}. This 
equation is easily generalized for 3D fermions and adopted to the case of a single boundary at 
$y=0$. It is convenient to present the tangential component of ${\bf u}$ as $u_{\alpha}(y) = 
u_h(y)+ \delta u(y)$, where hydrodynamic part $u_h$ obeys Eq. (7) at $y \gg l$ and 
$\delta u$ is a near-boundary correction. In the hydrodynamic regime, 
$u_h$ varies on the diffusion length $\sqrt{ll_1/(d+2)}$ which is much larger than $l$. 
Thus, near the boundary one can apply the linear form $u_h(y)= h+ h'y$ 
and reduce the integral equation for $u_{\alpha}(y)$ to the following one:
\begin{eqnarray}
\delta u(y) - \int_{0}^{\infty} \!\! \frac{d y'}{l} {\cal K}(y,y') \delta u(y') 
=l h'F^+_1(y)- hF^-_0(y),
\end{eqnarray}
where 
\begin{eqnarray}
{\cal K}(y,y')= a_d \int_0^{\pi/2} d \varphi \frac{\cos^d \varphi}{\sin \varphi}
\left[ e^{-|y-y'|/l \sin \varphi} \right. \nonumber \\ 
\left. + {\rm r}_{\varphi} e^{-(y+y')/ l \sin \varphi} \right], \\
F^{\pm}_k(y)=a_d \int_0^{\pi/2} d \varphi \cos^d \varphi \sin^k \varphi (1\pm {\rm r}_{\varphi}) 
e^{-y/ l \sin \varphi},
\end{eqnarray}
with $a_3=3/4$ and $a_2=2/\pi$. The approximate slip length given by Eq. (12) is found 
by integrating both sides of Eq. (13) over $y$ from $0$ to $\infty$ and neglecting the 
integral term in the resulting equation. Indeed, in this way one obtains $l_S=
l F^{+}_2(0)/F^{-}_1(0)$, which is identical to Eq. (12) at $B=0$ since $F^{\pm}_k(0)=
a_d \lambda^{\pm}_k$. The exact slip length $l_S=h/h'$ is determined from the requirement 
that the solution $\delta u(y)$ of Eq. (13) goes to zero at $y \gg l$. 
The difference $\delta l_S$ between the exact and the approximate slip lengths is
$\delta l_S= - \int_0^{\infty} dy F_0^{-}(y) \delta u (y)/l h' F^-_1(0)$. 
Whereas $\delta l_S$ can be either positive or negative, the relative deviation 
$\delta l_S/l_S$ is always numerically small. 

A comparison of the exact and approximate $l_S$ calculated within several 
models of ${\rm r}_{\varphi}$ is shown in Fig. 1 for both 3D and 2D systems. The 
model of angle-independent ${\rm r}_{\varphi}={\rm r}$ with ${\rm r} \in [0,1]$ is 
often applied in transport problems, though the absence of angular dependence
of the reflection probability is a rough assumption, especially near 
$\varphi=0$ and $\varphi=\pi$. The model ${\rm r}_{\varphi}=1 - (1-\beta) \sin \varphi$ 
with $\beta \in [0,1]$ reflects the property $P_{\varepsilon}({\bf p},{\bf p}') 
\propto p_n p'_n$ following from the treatment of boundary scattering in the Born 
approximation \cite{falkovski}. This model, however, does not describe fully 
diffuse boundaries. The model ${\rm r}_{\varphi}=\exp[-(\gamma^{-1}-1) \sin^2 \varphi]$ 
with $\gamma \in [0,1]$ is free from this disadvantage.

\begin{figure}[ht!]
\includegraphics[width=8.5cm,clip=]{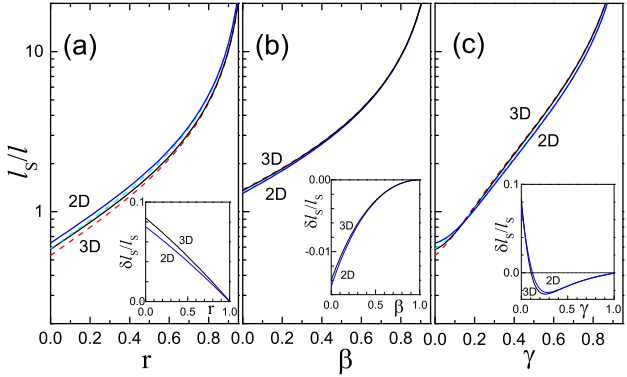}
\caption{\label{fig.1}(Color online) Exact (solid) and approximate (dashed) 
slip length for different specularity parameters: ${\rm r}_{\varphi}={\rm r}$ (a), 
${\rm r}_{\varphi}=1 - (1-\beta) \sin \varphi$ (b), and ${\rm r}_{\varphi}=
\exp[-(\gamma^{-1}-1) \sin^2 \varphi]$ (c). The insets show the relative 
deviation of the exact results from the approximate ones.}  
\end{figure} 

The slip length is the smallest for fully diffuse reflection, when the calculations 
give $l_S/l=0.582$ for 3D, in agreement with \cite{einzel2,latyshev}, 
and $l_S/l=0.637$ for 2D fermions. The absolute value of the relative deviation 
$\delta l_S/l_S$ is the largest in this case (however, less than 9\%), and decreases 
with increasing specularity. For angle-independent model, the exact 
$l_S$ does not simply scale as $(1+{\rm r})/(1-{\rm r})$ with respect to its value at 
${\rm r}=0$, but approaches the approximate value given by Eq. (12). For 
${\rm r}_{\varphi}=1 - (1-\beta) \sin \varphi$, both $|\delta l_S|/l_S$ and the difference 
between 2D and 3D cases are very small. The model ${\rm r}_{\varphi}=\exp[-(\gamma^{-1}-1) 
\sin^2 \varphi]$ shows sign inversion of $\delta l_S$: the approximate value of $l_S$, 
which is the lower bound of the exact solution for low specularity, becomes the upper 
bound for higher specularity at $\gamma \simeq 0.12$. 

\begin{figure}[ht!]
\includegraphics[width=8.5cm,clip=]{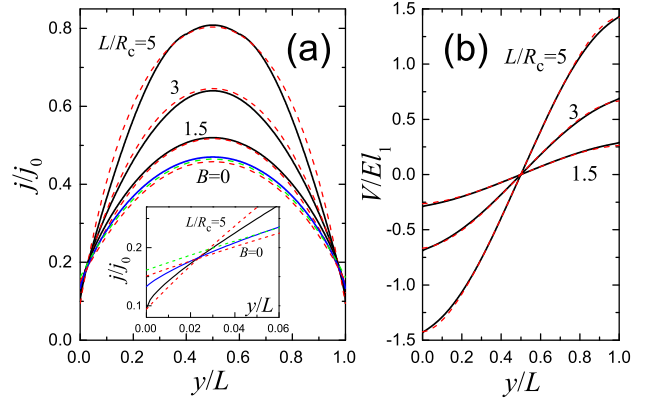}
\caption{\label{fig.2}(Color online) Kinetic (bold lines) and hydrodynamic (dashed red 
lines) distributions of the tangential current density $j(y)$ (a) and Hall potential $V(y)$ (b) 
across the 2D channel with equal fully diffuse boundaries at $l_{1}/L=5$ and 
$l_{1}/l_e=25$, for several values of $B$ defined by the ratio of channel width to 
cyclotron radius: $L/R_c=$0, 1.5, 3, and 5. The dashed green line is the distribution 
at $B=0$ obtained with the exact slip length $0.637 l$. The current and the 
potential are given in units of $j_0=\sigma_0 E$ and $El_1$, where $E$ 
is the driving electric field along the channel and $\sigma_0=e^2 n \tau_1/m$ is the 
Drude conductivity. The inset shows the region near $y=0$.}
\end{figure} 

\begin{figure}[ht!]
\includegraphics[width=8.5cm,clip=]{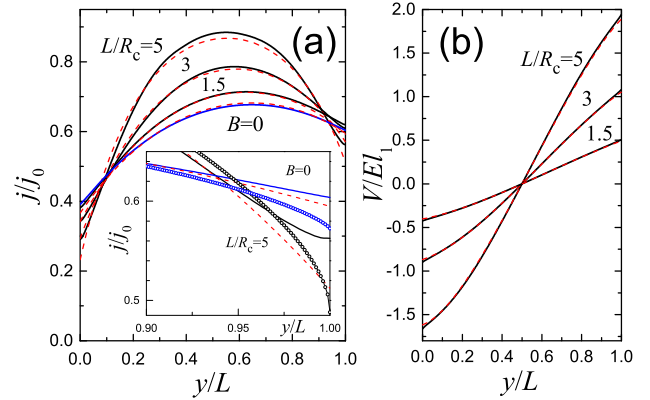}
\caption{\label{fig.3}(Color online) The same as in Fig. 2 for the channel 
with non-equal partly diffuse boundaries described by ${\rm r}_{\varphi}=
\exp[-(\gamma^{-1}-1)\sin^2 \varphi]$ with $\gamma=1/3$ ($l_S/l \simeq 1.7$) 
at $y=0$ and $\gamma=2/3$ ($l_S/l \simeq 6.0$) at $y=L$. The inset 
shows the region near $y=L$ and additional plots (circles) for the 
model of angle-independent specularity parameter with ${\rm r}=0.49$ ($l_S/l \simeq 1.7$) 
at $y=0$ and ${\rm r}=0.82$ ($l_S/l \simeq 6.0$) at $y=L$.}
\end{figure} 

The smallness of $\delta l_S/l_S$ justifies application of Eq. (12) 
in the hydrodynamic BC (1). This conclusion has been tested by calculating the 
distributions of current density $j(y)$ and electrochemical potential 
$V(y)$ in a narrow 2D channel of width $L$. The results obtained from 
a numerical solution of the kinetic equation with the BC (3) \cite{raichev1}
have been compared to the approximate analytical results obtained from Eq. (7) 
with the BC given by Eqs. (1) and (12), under the conditions of hydrodynamic 
regime, $l_e/l_1 \ll 1$ and $l_e/L \ll 1$. Several representative plots are shown 
in Figs. 2 and 3. If $l_e$ is estimated according to $l_e \simeq v 
\hbar \varepsilon_F/T^2$, the chosen ratios $l_{1}/L=5$ and $l_{1}/l_e=25$ can 
be achieved in GaAs 2D channels with $L \simeq 1.5$ $\mu$m at the densities $n 
\simeq 5 \times 10^{11}$ cm$^{-2}$ and temperatures $T \simeq 40$ K.
For equal boundaries, typical profiles of $j(y)$ and $V(y)$ \cite{sulpizio,holder,raichev1} 
are realized (Fig. 2), whereas the boundaries of different specularities lead to 
asymmetric profiles (Fig. 3). In both cases, the hydrodynamic approximation shows 
good agreement with the numerical solution at $B=0$. For highly diffuse boundaries 
(Fig. 2), the agreement is improved by using the exact $l_S$. Within the 
narrow Knudsen layers near the boundaries, the difference between the hydrodynamic and 
kinetic solutions is maximal, and the kinetic solution becomes most sensitive to 
the choice of specularity parameter model (Fig. 3, inset). The sign of this difference 
correlates with the sign of $\delta l_S$. 
Good agreement is also found at finite $B$, though the slopes of $j(y)$
expectedly differ near the boundaries within the cyclotron diameter $2R_c$ which 
defines the effective Knudsen layer at $R_c<l$. A detailed discussion 
of the underlying physics requires a solution of the problem of exact slip length in the 
presence of magnetic field, which is the subject for a future study. 

To summarize, a direct link between the BC for the distribution function and 
Maxwell's BC for the drift velocity is given by the approximate expression of the 
boundary slip length $l_S$ through the specularity parameter characterizing properties 
of the boundary scattering [Eqs. (10)-(12)]. For a gas of charged fermions in magnetic 
field $B$, $l_S$ is $B$-dependent and anisotropic, reflecting the behavior 
of the viscosity. The exact $l_S$, calculated at $B=0$ for several 
models of angle-dependent specularity parameter, is within a few percent from the 
approximate one and converges to it with increasing specularity [Fig. 1]. This finding, 
together with applications of the results to Poiseuille flow [Figs. 2 and 3], proves 
that the approximate expression for $l_S$ works much better than one might initially expect.

\end{document}